\documentclass[conference]{IEEEtran}
\IEEEoverridecommandlockouts
\usepackage{longtable}
\usepackage[noadjust]{cite}
\usepackage{amsmath,amssymb,amsfonts}
\usepackage{algorithmic}
\usepackage{graphicx}

\usepackage{comment}
\usepackage{textcomp}
\usepackage{xcolor}
\usepackage{adjustbox}
\usepackage{textcomp}
\usepackage[numbered]{bookmark}
\usepackage{soul}
\usepackage{booktabs}
\usepackage{multirow}
\usepackage{float}
\usepackage{tcolorbox}
\usepackage{graphicx}
\usepackage{url}
\usepackage{tabularx} 
\usepackage[justification=centering]{caption}
\usepackage{caption}
\usepackage{pdflscape}
\usepackage{subcaption}
\usepackage{rotating}
\usepackage[switch]{lineno}
\usepackage{ragged2e}
\usepackage{dirtytalk}
\usepackage{multicol}
\usepackage{csquotes}
\usepackage{pgfplots, pgfplotstable}
\usepackage{pgf-pie}
\usepackage{colortbl}
\usepackage{xspace}

\usepackage{array}
\usepackage{balance}
\usepackage{supertabular} 
\usepackage{fontawesome}
\usepackage[flushleft]{threeparttable}
\usepackage{supertabular}
\DeclareCaptionType{TextBox}
\newcolumntype{L}{>{\arraybackslash}m{16cm}}
\newcolumntype{C}[1]{>{\centering\let\newline\\arraybackslash\hspace{0pt}}m{#1}}
\newcolumntype{R}[1]{>{\raggedleft\let\newline\\arraybackslash\hspace{0pt}}m{#1}}
\def\BibTeX{{\rm B\kern-.05em{\sc i\kern-.025em b}\kern-.08em
    T\kern-.1667em\lower.7ex\hbox{E}\kern-.125emX}}
\begin{document}

\title{\huge State of Refactoring Adoption: Better Understanding Developer Perception of Refactoring\\
}

\author{\IEEEauthorblockN{Eman Abdullah AlOmar}
\textit{Stevens Institute of Technology}\\
Hoboken, NJ, USA \\
ealomar@stevens.edu}


\maketitle

\begin{abstract}

We aim to explore how developers document their refactoring activities during the software life cycle. We call such activity Self-Affirmed Refactoring (SAR), which indicates developers' documentation of their refactoring activities. SAR is crucial in understanding various aspects of refactoring, including the motivation, procedure, and consequences of the performed code change. After that, we propose an approach to identify whether a commit describes developer-related refactoring events to classify them according to the refactoring common quality improvement categories. To complement this goal, we aim to reveal insights into how reviewers decide to accept or reject a submitted refactoring request and what makes such a review challenging. 


Our SAR taxonomy and model can work with refactoring detectors to report any early inconsistency between refactoring types and their documentation. They can serve as a solid background for various empirical investigations.  Our survey with code reviewers has revealed several difficulties related to understanding the refactoring intent and implications on the functional and non-functional aspects of the software. In light of our findings from the industrial case study, we recommended a procedure to properly document refactoring activities, 
as part of our survey feedback. 

\end{abstract}


\section{Introduction}
\label{sec:Introduction}
The success of a software system depends on its ability to retain high-quality design in the face of continuous change. However, managing the growth of the software while continuously developing its functionalities is challenging and can account for up to 75\% of the total development \cite{erlikh2000leveraging,barry1981software}. 
 One key practice to cope with this challenge is refactoring. Refactoring is the art of remodeling the software design without altering its functionalities \cite{Fowler:1999:RID:311424}. It was popularized by Fowler \cite{Fowler:1999:RID:311424}, who identified 72 refactoring types and provided examples of how to apply them in his catalog.

Refactoring is a critical software maintenance activity that is  performed by developers for an amalgamation of reasons \cite{Tsantalis:2013:MES:2555523.2555539,Silva:2016:WWR:2950290.2950305,palomba2017exploratory}. Refactoring activities in the source code can be automatically detected \cite{Dig2006,Tsantalis:2013:MES:2555523.2555539} providing a unique opportunity to practitioners and researchers to analyze how developers maintain their code during different phases of the development life-cycle and over large periods of time. Such valuable knowledge is vital for understanding more about the maintenance phase; the most costly phase in software development \cite{Boehm:2002:SEE:944331.944370,erlikh2000leveraging}. To detect refactorings, the state-of-the-art techniques \cite{Dig2006,Tsantalis:2013:MES:2555523.2555539} typically search at the level of commits. As a result, these techniques are also able to group commit messages with their corresponding refactorings.

There are many opportunities to use refactoring to remove code smells. Code smell is the design defect that might violate the fundamentals of software design principles and decrease code quality \cite{Fowler:1999:RID:311424}. Examples of these code smells include duplicate code, dead code, long method, blob class, etc. An effective refactoring strategy that might be used to remove code duplication is to pull out the corresponding code element and locate it so that it can be shared in both objects resulting in extensibility, readability, etc. Contrastingly from a bug, a code smell does not necessarily cause a
fault or error in the application but may lead to other negative consequences, impacting software maintenance and quality.


\section{Problem Statement}
While existing studies have focused on automating the act of refactoring, refactoring tools are still underused as there is a lack of trust in these tools, and so, developers prefer to perform the manual refactoring. Additionally, recent studies questioned the quality metrics that we have used so far. These challenges reveal a lack of refactoring \textit{culture}. In this paper, we focus on the following challenges:
\begin{itemize}
    \item Commit messages are the description, in natural language, of the code-level changes. To understand the nature of the change, recent studies have been using natural language processing to process commit messages for multiple reasons, such as classification of code changes, change summarization, change bug-proneness, and developers' rationale behind their coding decisions. That is, commit messages are a common way for researchers to study developer rationale behind different types of changes to the code. There are two primary challenges to using commit messages to understand refactorings: 1) the commit message does not have to refer to the refactoring that took place at all, 2) developers have many ways of describing the same activity. For example, instead of explicitly stating that they are refactoring, a developer may instead state that they are \textit{performing code clean-up} or \textit{simplifying a method}. Developers are inconsistent in the way they discuss refactorings in commit messages. This makes it difficult to perform analysis on commit messages, since researchers may find it challenging to determine whether a commit message discusses the refactoring(s) being performed or not. Thus, it is hard to determine when the commit message is discussing a refactoring at all, and it is hard to determine how a commit message is discussing the refactoring. As the accuracy of refactoring detectors has reached a relatively high rate, the mined commits represent a
rich space to understand how developers describe, in natural
language, and their refactoring activities. Yet, such information
retrieval can be challenging since there are no common
standards on how developers should be formally documenting
their refactorings, besides inheriting all the challenges related
to natural language processing.
However, using the developer's inline documentation has
added another dimension to better understanding software
quality, as mining developers comments, for instance, has
unveiled how developers knowingly commit code that is either
incomplete, temporary, error-prone.
    \item Documenting refactoring,
similarly to any code change documentation, is helpful to decipher the
the rationale behind any applied change, and it can help future developers in various
engineering tasks, such as program comprehension, design reverse-engineering,
and debugging. The detection of such refactoring documentation was
hardly manual and limited. There is a need to automate the detection of
such documentation activities with an acceptable level of accuracy. Indeed, the
automated detection of refactoring documentation may support various applications
and provide actionable insights to software practitioners and researchers,
including empirical studies around the developer's perception of refactoring.
This can question whether developers care about structural metrics and
code smells when refactoring their code, or if other factors may
influence such non-functional changes.
    \item Despite the growing effort in recommending refactorings
through structural metrics, optimization and code smells removal,
there is very little evidence on whether developers
follow that intention when refactoring their code. Thus, there is a need to distinguish among all the structural
metrics, typically used in refactoring literature, the particular
ones that are a better representation of the developers’
perception of software quality improvement.
    \item Refactoring, just like any code change, has to be reviewed before being merged into the code base. However, little is known about how developers \textit{perceive}  and \textit{practice} refactoring during the code review process, especially since refactoring, by definition, is not intended to alter the system's behavior, but to improve its structure, so its review may differ from other code changes. Yet, there is not much research investigating how developers to review code refactoring. The refactoring research has focused on its automation by identifying refactoring opportunities in the source code and recommending adequate refactoring operations to perform. Moreover, the research on code reviews has been focused on automating them by recommending the most appropriate reviewer for a given code change. However, despite the critical role of refactoring and code review, their innate relationship is still largely unexplored in practice. 

\end{itemize}

\section{Research Goals}
To cope with the above-mentioned challenges, throughout this project, we aim to achieve the following research goals:
\begin{itemize}
    \item \textbf{Goal \#1:} \textit{Exploring how developers document refactoring activities.} We aim to extract how developers express
their nonfunctional activities, namely improving software design,
renaming semantically ambiguous identifiers, removing
code redundancies etc. Multiple studies have been detecting
the performed refactoring operations, \textit{e.g.,} rename class, move
method etc. within committed changes to better understand
how developers cope with bad design decisions, also known
as design antipatterns, and to extract their removal strategy
through the selection of the appropriate set of refactoring
operations \cite{tsantalis2018accurate}. 
    
    \item \textbf{Goal \#2:} \textit{Understanding developer perception of refactoring.} We aim to
augment our understanding of the development contexts that trigger refactoring
activities and enable future research to take development contexts into account
more effectively when studying refactorings. Thus, the advantages of analyzing
the textual description of the code change that was intended to describe
refactoring activities are three-fold: 1) it improves our ability to study and commit
message content and relate this content to refactorings; a challenging task which
posed a significant hurdle in recent work on contextualizing rename refactorings, 2) it gives us a stronger understanding of commit
message practices and could help us improve commit message generation by
making it clear how developers prefer to express their refactoring activities, 3) it
provides us with a way of relating common words and phrases used to describe
refactorings with one another. We plan to progress on refactoring motivation direction by identifying, among the various quality
models presented in the literature, the ones that are more inline
with the developer’s vision of quality, when they explicitly
state that they are refactoring the code to improve it. 

 \item \textbf{Goal \#3:} \textit{Studying developers refactoring perspective in practice.} We aim to survey professional developers and conduct a case study in the industry to gain practical insights from refactoring in practice by studying refactoring motivation, documentation practices, and challenges. This direction of study can help us design future studies in refactoring that are empirically relevant to practitioners' obstacles, challenges, and needs, and create the next generation of industry-relevant automated refactoring tools.
 
    
\end{itemize}

\section{Contributions}

\begin{figure*}
\centering 
\includegraphics[width=\textwidth]{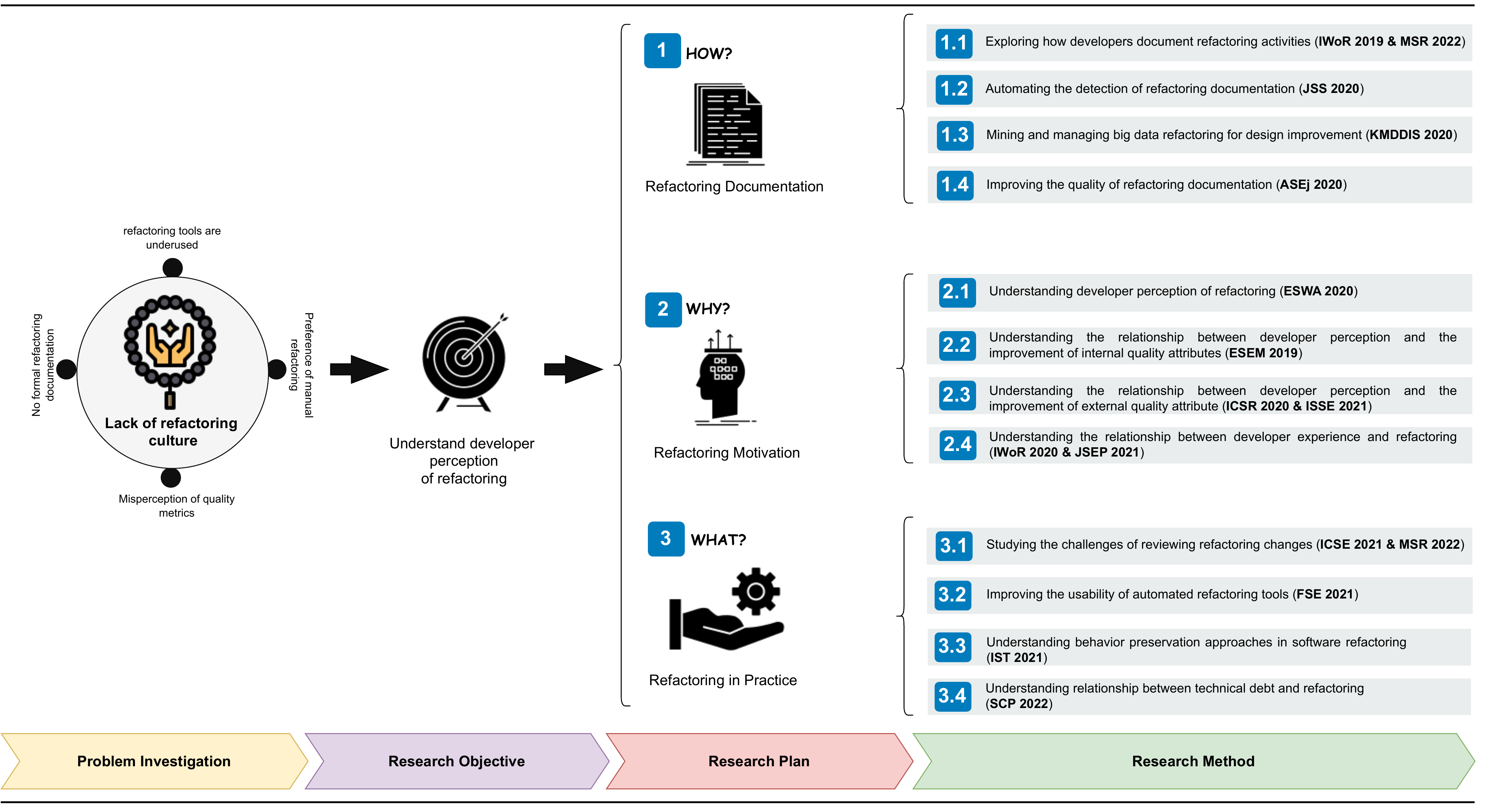}
\caption{Overview of the contributions.}
\label{fig:thesiscontribution}
\end{figure*}

In accomplishing the goals, the
solutions are organized into 3 main contributions as it is shown in Figure \ref{fig:thesiscontribution}.

\begin{itemize}

   \item In our research (IWoR2019 \cite{alomar2019can}, JSS2020 \cite{alomar2020toward}, KMDDIS2020 \cite{alomarmining}, ASEj2021 \cite{AlOmar2021ondocumentation}, MSR2022 \cite{alomar2022exploratory}), we extracted how developers express
their nonfunctional activities, namely improving software design,
renaming semantically ambiguous identifiers, removing
code redundancies etc. 

     \item In our research (ESEM2019 \cite{alomar2019impact}, ICSR2020 \cite{alomar2020how}, ESWA2020 \cite{alomar2021we}, IWoR2020 \cite{AlOmarIWoR2020}, JSEP2021 \cite{AlOmar2021behind}, ISSE2021 \cite{AlOmar2021refactoringreuse}), we
augmented our understanding of the development contexts that trigger refactoring
activities and enable future research to take development contexts into account
more effectively when studying refactorings. 

\item In our research (ICSE2021 \cite{alomar2021refactoring}, MSR2022 \cite{alomar2022code}, FSE2021 \cite{Golubev2021FSE}, IST2021 \cite{alomar2021SLM}, SCP2022 \cite{alomar2022satdbailiff}), we surveyed professional developers and conduct a case study in the industry to gain practical insights from refactoring in practice by studying refactoring motivation, documentation practices, and challenges. 

\end{itemize}

\noindent\textbf{Research repositories:} The comprehensive experiments package are available online in \cite{SAR2020WEB}.

\section{Research Implication}

\faThumbTack \noindent{\textbf{ Augmenting refactoring automation with documentation.} Recent studies have been focusing on automatically identifying any execution of a refactoring operation in the source code \cite{silva2017refdiff,tsantalis2018accurate}. The main purpose of the automatic detection of refactoring is to understand better how developers cope with their software decay by extracting any refactoring strategies that can be associated with removing code smells \cite{tsantalis2008jdeodorant,bavota2013empirical}, or improving the structural design measurements \cite{bavota2014recommending}. However, these techniques only analyze the changes at the source code level, and provide the operations performed, without associating it with any textual description, which may infer the rationale behind the refactoring application. Our proposed model intends to bridge this gap by complementing the existing effort in accurately detecting refactorings, by augmenting with any description that was intended to describe the refactoring activity. As per our findings, developers tend to add a high-level description of their refactoring activity, and occasionally mention their intention behind refactoring (remove duplicate code, improve readability), along with mentioning the refactoring operations they apply (type migration, inline methods, etc.). Our model, combined with detecting refactoring operations, serves as a solid background for various empirical investigations. For instance, previous studies have analyzed the impact of refactoring operations on structural metrics \cite{cedrim2016does}. One of the main limitations of these studies is the absence of any context related to the application of refactorings, \textit{i.e.,} it is not clear whether developers did apply these refactoring with the intention of improving design metrics. Therefore, using our model will allow the consideration of commits whose commit messages specifically express the refactoring to optimize structural metrics, such as coupling and complexity. So, many empirical studies can be revisited with an adequate dataset. 

Furthermore, our study provides software practitioners with a catalog of common refactoring documentation patterns  which would represent concrete examples of common ways to document refactoring activities in commit messages. This catalog of SAR patterns can encourage developers follow best documentation patterns and also to further extend these patterns to improve refactoring changes documentation in particular and code changes in general. Indeed, reliable and accurate documentation is crucial in any software project. The presence of documentation for low-level changes such as refactoring operations and commit changes helps to keep track of all aspects of software development, and it improves the quality of the end product. Its main focuses are learning and knowledge transfer to other developers.

\faThumbTack \noindent{\textbf{ Understanding developer's motivation behind refactoring.} One of the main findings shows that developers are not only driven by design improvement and code smell removal when taking decisions about refactoring. According to our findings, fixing bugs, and feature implementation play a major role in triggering various refactoring activities. Traditional refactoring tools are still leading their refactoring effort based on how it is needed to cope with design antipatterns, which is acceptable to the extent where it is indeed the developer's intention, otherwise, they have not been designed or tested in different circumstances. So, an interesting future direction is to study how we can augment existing refactoring tools to better frame the developer's perception of refactoring, and then their corresponding objectives to achieve (reducing coupling, improving code readability, renaming to remove ambiguity etc.). This will automatically induce the search for more adequate refactoring operations, to achieve each objective.

The categories of refactoring motivation provide software practitioners with a catalog of common documentation patterns that represent concrete examples of common ways to document refactoring activities in commit messages. Having these higher-level categories helps developers find the specific refactoring patterns they are looking for faster. Generally, in industry, there is no guideline on structuring commit messages. This catalog of SAR patterns can encourage developers to follow the best documentation patterns and further extend these patterns to improve refactoring documentation in particular and code changes in general. This work will also help developers to improve the quality of the refactoring documentation and trigger the need to explore the motivation behind refactoring. Further, these categories tell the opinion of developers, so it is essential for managers to learn developers' opinions and feelings, especially for distributed software development practices. If developers do not document, managers will not know their intention. Since software engineering is a human-centric process, it is essential for managers to understand the intention of people working on the team.  

\faThumbTack \noindent{\textbf{ Understanding code review practice for refactoring changes.} It is heartening for us to realize that developers refactor their code and perform reviews for the refactored code. Our main observation, from developers' responses, is how the review process for refactoring is being hindered by the lack of documentation. Therefore, as part of our survey report to the company, we designed a procedure for documenting any refactoring ReR, respecting three dimensions that we refer to as the three \textit{\textbf{I}}s, namely, \textit{\textbf{I}ntent}, \textit{\textbf{I}nstruction}, and \textit{\textbf{I}mpact}. 
 We detail each one of these dimensions as follows:

\textbf{Intent.} According to our survey results, it is intuitive that reviewers need to understand the purpose of the intended refactoring as part of evaluating its relevance. Therefore, when preparing the request for review, developers need to start with explicitly stating the motivation of the refactoring. This will provide the context of the proposed changes, 
for the reviewers, so they can quickly identify how they can comprehend it. According to our initial investigations, examples of refactoring intents include \textit{enforcing best practices}, \textit{removing legacy code}, \textit{improving readability}, \textit{optimizing for performance}, \textit{code clean up}, and \textit{splitting logic}.

\textbf{Instruction.} Our results show how rarely developers report refactoring operations as part of their documentation. Developers need to clearly report all the refactoring operations they have performed, in order to allow their reproducibility by the reviewers. Each instruction needs to state the type of the refactoring (move, extract, rename, etc.) along with the code element being refactored (\textit{i.e.,} package, class, method, etc.), and the results of the refactoring (the new location of a method, the newly extracted class, the new name of an identifier, etc.). If developers have applied batch or composite refactorings, they need to be broken down for the reviewers. Also, in case of multiple refactorings applied, they need to be reported in their execution chronological order.

\textbf{Impact.} We observe that practitioners care about understanding the impact of applied refactoring. Thus, the third dimension of the documentation is the need to describe how developers ensure that they have correctly implemented their refactoring and how they verified the achievement of their intent. For instance, if this refactoring was part of a bug fix, developers must reference the patch. If developers have added or updated the selected unit tests, they need to attach them as part of the review request. Also, it is critical to self-assess the proposed changes using Quality Gate, to report all the variations in the structural measurements and metrics (\textit{e.g.,} coupling, complexity, cohesion, etc.), and provide a necessary explanation in case the proposed changes do not optimize the quality deficit index.

\faThumbTack \noindent{\textbf{ Developing next-generation refactoring-related code review tools.}  Finding that reviewing refactoring changes takes longer than non-refactoring changes reaffirms the necessity of developing accurate and efficient tools and techniques that can assist developers in the review process in the presence of refactorings. The refactoring toolset should be treated like the CI/CD toolset and integrated into the tool chain. 
 Researchers could use our findings with other empirical investigations of refactoring to define, validate, and develop a scheme to build automated assistance for reviewing refactoring considering the refactoring review criteria as review code become an easier process if the code review dashboard augmented with the factors to offer suggestions to document the review better.

\faThumbTack \noindent{\textbf{ Teaching Documentation Best Practices.} Prospective software engineers are mainly taught how to model, develop and maintain software. With the growth of software communities and their organizational and socio-technical issues, it is important also to teach the next generation of software engineers the best practices of refactoring documentation. So far, these skills can only be acquired by experience or training. 

\bibliographystyle{ieeetr}
{\scriptsize\bibliography{references}}

\end{document}